# Comment on „Improper molecular ferroelectrics with simultaneous ultrahigh pyroelectricity and figures of merit" by Li *et al.*

Marek Szafrański[1†], Andrzej Katrusiak[2†]

[1]Faculty of Physics, Adam Mickiewicz University, Uniwersytetu Poznańskiego 2, 61-614 Poznań, Poland;
[2]Faculty of Chemistry, Adam Mickiewicz University, Uniwersytetu Poznańskiego 8, 61-614 Poznań, Poland;

[†]Corresponding author. Email: masza@amu.edu.pl (M.S.); katran@amu.edu.pl (A.K)

**Abstract**

Li *et al.* (*Science Advances*, 29 January, p. eabe3068) claim the discovery of two improper ferroelectrics, dabcoHClO$_4$ and dabcoHBF$_4$ (dabco = 1,4-diazabicyclo[2.2.2]octane), and that these materials exhibit superior pyroelectric figures of merit. This information is misleading due to the fundamental methodological errors and false conclusions, not to mention that these ferroelectrics were reported over 20 years ago. They are proper ferroelectrics, for which the spontaneous polarization is the macroscopic order parameter. We show that the useful pyroelectric coefficients of these materials are about $10^3$ times lower than these reported by Li *et al.*

In a recently published article, Li *et al.* (*1*) report the ferroelectricity of two hybrid organic-inorganic crystals, dabcoHClO$_4$ and dabcoHBF$_4$ (dabco = 1,4-diazabicyclo[2.2.2]octane, C$_6$H$_{12}$N$_2$), and claim that these materials exhibit superior pyroelectric properties related to the improper nature of ferroelectricity. In fact, this is a second rediscovery of ferroelectric properties in these NH⋯N bonded crystals, after the first rediscovery published by Xiong's group in 2016 (*2*). Both these papers (*1,2*) carefully conceal the original report on ferroelectricity in dabco monosalts (*3*) published in 1999. Li *et al.* mention our report in page 5 of their paper, not in the connection with ferroelectricity of dabcoHClO$_4$ and dabcoHBF$_4$, but incorrectly referring to "improper ferroelectricity in molecular materials (*27-29*)", whereas there is no such information in our paper. Leaving aside the ethical issues, the paper by Li *et al.* contains many serious errors and discrepancies, which require to be commented and corrected.

DabcoHClO$_4$ and dabcoHBF$_4$ can be described as ionic, molecular ionic or hybrid organic-inorganic materials, but certainly not as the molecular ones. The structural and symmetry information is essential for understanding the mechanism underlaying their ferroelectric properties and phase transitions. Therefore, we have performed over 50 single-crystal X-ray and neutron diffraction experiments (*3-6*), as a function of temperature and under pressure. The single-crystal measurements circumvent the overlapping of reflections inherent to powder X-ray diffraction (PXRD), hampering the precise structural determinations and location of the disordered atoms and protons. All our structural determinations were deposited in the Cambridge Structural Database and can be received free of charge. Li *et al.* do not mention these CSD deposits and claim that they determined the structures of the ferroelectric and highly disordered paraelectric phases, from in-house PXRD measurements. They describe incorrectly the symmetry of paraelectric phases of dabcoHClO$_4$ and dabcoHBF$_4$ as space group *P*4/*mmm* (No 123) with the unit cell containing one ionic pair (Z=1), while, as shown in our previous study (*4*), the true space group is *P*4/*nmm* (No 129) and the unit cell is two times larger (Z=2). Despite the wrong space group and lattice parameters, the structural model presented by Li *et al.* in Figs. 1D and 1F is consistent with the correct lager unit cell and symmetry determined in our single-crystal experiments (*4*). The symmetry relation between the ferroelectric and paraelectric phases is of primary importance. For the symmetry change *Pm*2$_1$*n* → *P*4/*mmm*, the translational symmetry of the crystals would be broken, suggesting a possible improper ferroelectricity (*7*), whereas the *Pm*2$_1$*n* → *P*4/*nmm* transition is clearly equi-translational, and accordingly dabcoHClO$_4$ and dabcoHBF$_4$ should be classified as proper ferroelectrics (*8*). Moreover, the improper ferroelectricity is associated with a minor change in the electric permittivity at $T_C$, and the Curie-Weiss law is not fulfilled in the paraelectric phase (*7,9*). Figs. 1A and 1B show the temperature dependence of electric permittivity, measured by us at 500 kHz for the single crystals of dabcoHBF$_4$ and dabcoHClO$_4$ along [010]. Clearly there are no features characteristic for improper ferroelectrics. On the contrary, the magnitude of the changes at $T_C$, the shape of the anomalies, and the fulfillment of the Curie-Weiss law in the paraelectric phases, fully reaffirm the proper ferroelectricity of both materials. Thus, dabcoHClO$_4$ and dabcoHBF$_4$ are neither molecular nor improper ferroelectrics, as claimed by

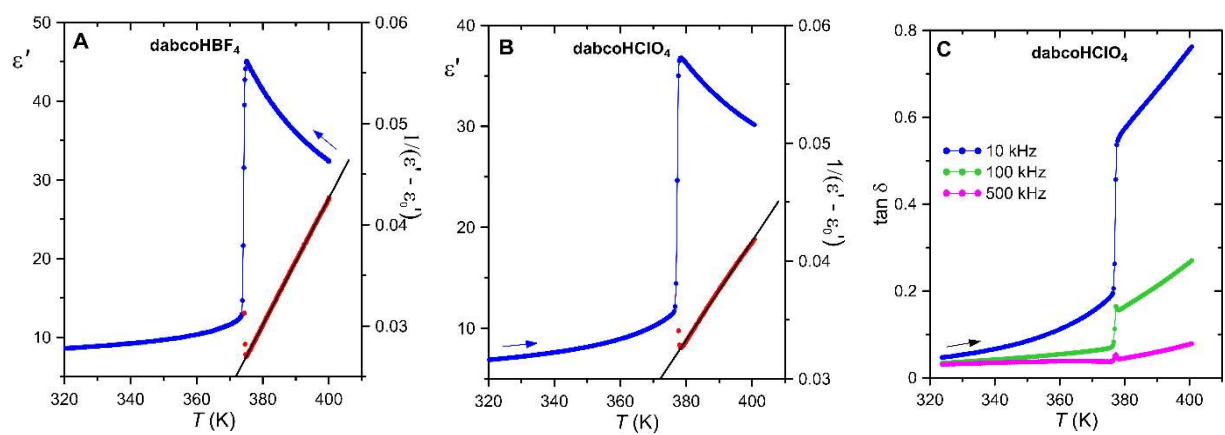

**Fig. 1. Dielectric response of single crystals of dabcoHBF$_4$ and dabcoHClO$_4$.** (**A**, **B**) The real part of electric permittivity, $\varepsilon'$, measured at 500 kHz along [010] for temperature changing at a rate 0.1 K/min (blue, left axes) and the illustration of the Curie-Weiss law fulfillment (red, right axes). (**C**) The low-frequency loss tangent, tan δ, for dabcoHClO$_4$ in the vicinity of the ferroelectric-paraelectric phase transition. Arrows indicate the directions of temperature changes.

Li *et al.* Their another doubtful information concerns the loss tangent (tan δ) of dabcoHClO$_4$ they determined at 1 kHz as 0.001 below and 0.08 above $T_C$ [Fig. 2A in (*1*)]. Such a low loss is exciting, but hardly possible for hybrid organic-inorganic H-bonded materials, exhibiting substantial dynamical disorder in the ferroelectric phases below $T_C$, and highly disordered above $T_C$. Indeed, the dielectric loss measured for the single crystal of dabcoHClO$_4$ in dry atmosphere (Fig. 1C) is more than one order of magnitude higher compared to that reported by Li *et al.* for the polycrystalline sample. The strong frequency dispersion and large increase of the loss at low frequencies are of primary importance for pyroelectric and polarization hysteresis loop measurements. The effect of electric conductivity on the hysteresis loop is evident in Fig. 2A at 346 K. We tested several single-crystal samples and each measurement proved impossible above c.a. 360 K, whereas Li *et al.* observed the ferroelectric polarization loop even at 383 K, i.e. in the paraelectric phase, 5 K above $T_C$ [Fig. 2B in (*1*)], where $P_s = 0$ according to their plot in Fig. 2C. A high value of $P_s = 6$ µC cm$^{-2}$ determined by Li *et al.* for the polycrystalline dabcoHClO$_4$ requires a comment, too. It is evident from the electric permittivity measurements presented in Fig. 2A in (*1*) that the crystalline grains were randomly oriented in the sample, as testified by 2.5 times lower permittivity in the paraelectric phase compared to our single-crystal results shown in Fig. 1B. Because $P_s$ cannot rotate in a ferroelectric crystal, only its components parallel to the applied electric field can contribute to the ferroelectric loop. Therefore, a much lower value of polarization is expected for polycrystalline sample than 4.6 µC cm$^{-2}$ determined from the single-crystal loop (Fig. 2A). The spontaneous polarization and the related pyroelectricity are the key issues reported by Li *et al*. Surprisingly, they determined the pyroelectric coefficient and the corresponding figures of merit [Figs. 2D-G in (*1*)] at $T_C$ of the first-order transition, and compare them with the values determined off the phase transitions for other pyroelectric materials [Fig. 3 in (*1*)]. This is a serious methodological error leading to false conclusions. The sharpness of the transition in dabcoHClO$_4$ can be rated from our plots in Figs. 1B and 2B. The $P_s(T)$ in Fig. 2B was measured on the virgin not poled crystal, which was spontaneously polarized, but most probably not completely, as most of the crystals grown. It is apparent that the transition occurs in a very narrow temperature interval $\Delta T < 1$ K, even for temperature changing at 2 K/min. By lowering this rate, the transition interval could be reduced to the limit $\Delta T \rightarrow 0$, and the pyroelectric coefficient $p = dP_s/dT$ would reach record values. It is clear that so determined $p$ at $T_C$ characterizes the transition process rather than the pyroelectric properties of the material. Moreover, it is not true that dabcoHClO$_4$ can be cycled across $T_C$ without a poling field and without a worsening of pyroelectric properties. This is evident from a decay of the polarization in the cooling run plotted in Fig. 2B. The $P_s(T)$ dependences in Fig. 2B represent the raw data, not corrected for the electric conductivity, while its effect is clearly seen in the paraelectric phase, where the nonzero pyroelectric charge (current) originates from the increased dielectric loss (see Fig. 1C). The practically useful temperature region, where the pyroelectric coefficient values are not nonrecurring but reproducible, is below $T_C$. However, as shown in Fig. 2C in this region $p$ does not exceeds $2.5 \times 10^{-2}$ µC cm$^{-2}$ K$^{-1}$ and hence dabcoHClO$_4$ in this respect is not superior when compared to the best pyroelectric materials.

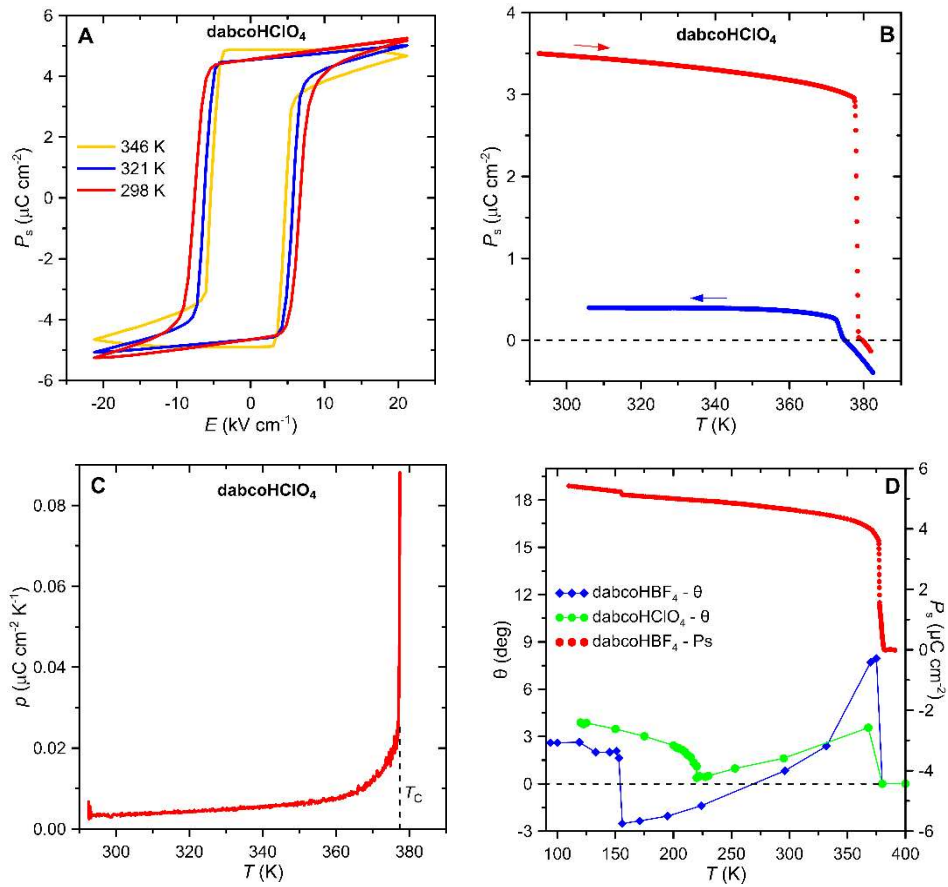

**Fig. 2. Spontaneous polarization and pyroelectric properties of dabcoHBF₄ and dabcoHClO₄.** (**A**) The low-frequency (0.1 Hz) ferroelectric hysteresis loops measured at different temperatures on a single-crystal dabcoHClO₄ along [010]. (**B**) Spontaneous polarization measured on as grown single crystal of dabcoHClO₄ along [010], during the heating (red) and subsequent cooling (blue) runs, at the rate of temperature changes 2 K/min. (**C**) Pyroelectric coefficient $p$ of dabcoHClO₄ determined from the heating run shown in (**B**). (**D**) Spontaneous polarization measured for single-crystal dabcoHBF₄ across the ferroelectric-paraelectric and ferroelectric-ferroelectric phase transitions according to (*6*) (right axis), and the $\theta$ angles (left axis) determined for dabcoHClO₄ and dabcoHBF₄ from our structural studies (*3-6*).

In the theoretical approach presented in (*1*) the authors try to apply the Landau-Devonshire theory to improper ferroelectrics. Disregarding the proper nature of ferroelectricity in dabco monosalts, their approach is disqualified by the incorrect choice of the order parameter, the dihedral angle $\theta$ between three oxygen/fluorine atoms of the anion and crystal plane (010)]. Levanyuk and Sannikov (*7*) showed that for improper ferroelectrics the one-component order parameter is inapplicable, because of the mixed terms $\theta P_s$ in the thermodynamic potential. The presence of invariant $\theta P_s$ in the free energy used by Li *et al.* [see their equation (1)] implies that $\theta$ and $P_s$ have the same transformation properties, and therefore the transition cannot be improper, as they assume. In Fig. 4A they compare $P_s(T)$ with their theoretical model, but it should be noted that the shape of the experimental $P_s(T)$ is distorted by the non-equilibrium conditions of the pyroelectric method, hence such a comparison is dubious. However, the most astonishing is the modelling of $P_s(T)$ in the phase transition region, above $T_C$, using a continuous function, whereas for the first-order phase transition $P_s$ should abruptly vanish at $T_C$.

The inadequacy of angle $\theta$ as the order parameter for dabco ferroelectrics is also evident from the plots in Figs. 2D and 2B. It is apparent that $\theta(T)$ and $P_s(T)$ are not related, while a tight coupling is predicted by the theoretical simulations illustrated in their Fig. 4B. This discrepancy could be expected, because $P_s$ in dabcoHClO$_4$ and dabcoHBF$_4$ originates from the ionic displacements δ, but not from the rotation of the tetrahedral anions. The structural origin of $P_s$ was described in our paper (*3*) and successfully used for $P_s(T)$ modelling (*3,6*). Thus, δ and $P_s$ are the true microscopic and macroscopic order parameters for describing the ferroelectric-to-paraelectric phase transitions in dabcoHClO$_4$ and dabcoHBF$_4$, as it was shown over 20 years ago.